\documentstyle[epsf,12pt]{article}
\let\section=\subsection \let\subsection=\subsubsection

\begin{document}
hep-ph/9703258 \hfill CALT-68-2104
\vskip0.2in

\begin{center}
{\large \bf DISORIENTING THE CHIRAL CONDENSATE}\\[2mm]
{\large \bf AT THE QCD PHASE TRANSITION}\footnote{Talk given at 
the International Workshop 
on QCD Phase Transitions held in January, 1997 in Hirschegg, Austria.}\\[5mm]
KRISHNA~RAJAGOPAL\\[5mm]
{\small \it Lauritsen Laboratory of High Energy Physics\\
California Institute of Technology, Pasadena, CA USA 91125\\[8mm] }
\end{center}

\begin{abstract}\noindent
I sketch how long wavelength modes of the pion field can be
amplified during the QCD phase transition. If nature had
been kinder, and had made the pion mass significantly less
than the critical temperature for the transition, then this
phenomenon would have characterized the transition in
thermal equilibrium.  Instead, these long wavelength oscillations
of the orientation of the chiral condensate can only arise
out of equilibrium.  There is a simple non-equilibrium
mechanism, plausibly operational during heavy ion collisions,
which naturally amplifies these oscillations. The characteristic
signature of this phenomenon is large fluctuations in the ratio
of the number of neutral pions to the total number of pions in
regions of momentum space, that is in phase space in a detector.
Detection in a heavy ion collision would imply an out of equilbrium
chiral transition.
\end{abstract}

Shortly after the discovery that QCD is asymptotically free \cite{asympfree}
and that therefore quarks are weakly interacting at short distances,
Collins and Perry \cite{collins} noted that this means that 
at temperatures 
$T\gg\Lambda_{\rm QCD}$,  the theory describes a world of weakly
interacting quarks and gluons very different from the familiar hadronic
world.\footnote{In this talk I will limit myself to discussing the
physics of a QCD plasma with zero baryon number density.
This should be a good approximation in
the central rapidity regions of 
heavy ion collisions at RHIC energies and higher, where
the number of pions per event is expected to be about 
two orders of magnitude
larger than the baryon number per event.} There are at least
two qualitative differences 
between a plasma at $T\gg \Lambda_{\rm QCD}$
and at $T\sim 0$. 
First, at low temperatures one has a plasma
of hadrons, while at high temperatures the quarks and gluons are
deconfined. Second, at low temperatures a $q\bar q$ condensate 
spontaneously breaks chiral symmetry while at high temperature 
the interactions among quarks and anti-quarks are weak, no such
condensate exists, and chiral symmetry is manifest. 

In $SU(3)$ gauge theory with no quarks, there are
analytic arguments\cite{svetitsky} confirmed by lattice
simulation\cite{pureglue} that the deconfinement transition
is first order.  While this conclusion presumably remains valid
if all quarks are much heavier than $\sim\Lambda_{\rm QCD}$,
no deconfinement order parameter  is known for $SU(N)$ 
gauge theory including dynamical quarks, and simulations
with two light quarks\cite{latt2ndorder,MILC}  find that 
deconfinement occurs via a smooth crossover.

The presence of light quarks
raises the possibility of
a phase transition associated with the vanishing of the
chiral order parameter.
The Lagrangian for QCD with two massless quarks has a global
$SU(2)_L\times SU(2)_R$ symmetry which, at low temperatures, 
is spontaneously broken to $SU(2)_{L+R}$ by a nonzero 
expectation value for the chiral order parameter, which
can be written in terms of four real scalar fields $(\sigma,\vec \pi)$
and the Pauli matrices
according to
\begin{equation}
\langle \bar q_L^i q_{Rj} \rangle = \sigma \delta^i_j + i \vec\pi \cdot 
\vec\tau^i_j \ .
\label{sigpidef}
\end{equation}
In fact, the order parameter can be written as a four component
scalar field
\begin{equation}
\phi^\alpha \equiv (\sigma, \vec\pi)\ ,
\label{phidef}
\end{equation}
and the $SU(2)_L\times SU(2)_R$ transformations are simply
$O(4)$ rotations.  At low temperatures, 
$\langle \phi \rangle$ is nonzero. This picks a direction
in $O(4)$ space defined to be the $\sigma$ direction,
and spontaneously breaks the symmetry.
The direction in which the
expectation value points is defined as the $\sigma$ direction.
Oscillations of 
$\phi$  
in the $\vec \pi$ directions, that is oscillations of the 
orientation of the condensate,
are massless Goldstone modes.
As $T$ is increased, $\phi$ fluctuates more and more wildly until
above some critical temperature $T_c$ the fluctuations are large enough
that $\langle \phi \rangle = 0$ and the $O(4)$ symmetry is restored.

Since the order parameter appropriate for the chiral phase transition
in QCD with two flavors of massless quarks has the symmetries of
an O(4) Heisenberg magnet, and since this model has a second order
phase transition,\footnote{This has been verified perturbatively
in the coupling, perturbatively in $d-4$ where $d$ is the number of
dimensions of space, perturbatively 
in $1/N$ where $N=4$ is the case of interest,
nonperturbatively in numerical simulations, and by experiment
for $N=3$.}
it is possible that the chiral phase transition
is second order and is in the same universality 
class.\cite{piswil,wil,RW1}  
At $T=T_c$ for a second order phase transition, the theory
is at an infrared fixed point of the renormalization group
and physics is scale invariant.
This means that the 
order parameter fluctuates on all length scales,
and in particular on arbitrarily long length scales.
Long wavelength oscillations of the chiral order 
parameter are the defining 
feature of physics near $T_c$, and our goal later in this talk
will be to discuss whether they may occur
in a heavy ion collison and how they can leave an observable
signature.     

If the transition is in the $O(4)$ universality class, 
the equilibrium physics of the order parameter 
near $T=T_c$ 
at wavelengths long compared to $1/T$ 
is classical\footnote{Equilibrium 
finite temperature quantum
field theory is formulated in 4 dimensional Euclidean space,
where the fourth dimension is periodic with period $1/T$.
Integrating out modes with energy $\sim 1/T$ and higher, 
yields an effective three dimensional classical field theory
describing physics at wavelengths longer than $1/T$.}
and is described by the Ginzburg-Landau free energy
\begin{equation}
F = \int d^3 x \left\{ \frac{1}{2}
\partial^i \phi^\alpha \partial_i \phi_\alpha
+ \frac{\mu^2}{2}\phi^\alpha\phi_\alpha + \frac{\lambda}{4} 
(\phi^\alpha \phi_\alpha)^2 + H \sigma \right\}\ .
\label{freeenergy}
\end{equation}
Here $\mu^2$ and $\lambda>0$ are temperature dependent and $T_c$ is
the temperature at which $\mu^2=0$. 
We have
introduced an explicit symmetry breaking term $H\sigma$ which
tilts the potential, 
selects a $\sigma$ direction, and gives the Goldstone
bosons a mass.  If the underlying microscopic
theory which under renormalization flows to (\ref{freeenergy}) 
in the infrared were in fact that of
a magnet, $H$ would be proportional to an externally imposed
magnetic field.  In QCD, $H$ is proportional to a common
mass $m_q$ for the two light quarks.  
A nonzero $H$
turns the second order phase transition into a smooth crossover.
If $H$ is small,  
the theory (\ref{freeenergy}) yields
universal, quantitative predictions for the 
behavior of the order parameter 
$\langle \phi \rangle$
as a function of $T$ for $T$ near 
$T_c$.\cite{wil,RW1,myreview,detar,toussaint}
The critical exponents describing physics at 
$H=0$, $T\neq T_c$ and at $T=T_c$, $H=0$ have been computed to
high order in a perturbative expansion in 
$\lambda$,\cite{bmn} and   
by numerical simulation.\cite{kk}
The full equation of state --- namely $\langle \phi \rangle$
as a function of $T$ and $H$ ---
has been obtained numerically by Toussaint.\cite{toussaint}  

Finite temperature lattice QCD simulations can
test whether the QCD phase transition is in fact in the 
$O(4)$ universality class.  
Present simulations provide
evidence that in two flavor QCD the phase transition is
not first order, and is plausibly second order
in the chiral limit.\cite{latt2ndorder,MILC}
As one example, let us consider
the results of \cite{MILC}.  
They compute
the order parameter as a function of $T$ over a range of temperatures
near $T_c$ for two values of $m_q$, the lighter of which corresponds
to $m_\pi/m_\rho \sim 0.3$.  They find that the order parameter
decreases rapidly but smoothly 
over a range of temperatures about $10 {\rm ~MeV}$ wide
centered about a temperature $T_c \sim 140 - 160 {\rm ~MeV}$.
They then show that their results 
can be fitted equally well using the $O(4)$ equation of state\cite{toussaint}
and the mean field equation of state.  However, the two fits make
very different predictions for 
behavior at smaller quark masses.  Thus, the present simulation
is consistent with the hypothesis that the transition is
in the $O(4)$ universality class, but until smaller quark masses
are explored no stringent tests will be possible.  
To see the
universal long wavelength physics one must use light enough quarks that
$m_\pi(T_c)<T_c$, where $m_\pi(T_c)$,
the pion inverse correlation length, 
is of order or slightly more than $m_\pi$.
In the simulations of \cite{MILC}, $m_\pi > T_c$.
This means that 
the correlation length is just not long enough for the physics
to become effectively three dimensional and classical as in
(\ref{freeenergy}).  
In present simulations
done by a number of groups\cite{latt2ndorder,MILC} which have all
had $m_\pi/m_\rho \sim 0.2$ or more,
although a variety of methods of
extracting critical exponents have been explored, it has not
yet been possible to cleanly distinguish between, say, $O(4)$ and 
mean field exponents.\footnote{It 
is worth enumerating the logical possibilities
which could lead to a failure of the hypothesis that a second order
phase transition with $O(4)$ exponents will be seen in lattice
simulations.  We only know that an infrared fixed point
with the appropriate symmetry exists. We do not know that
QCD is in fact in the basin of attraction of this fixed point.
Renormalization toward the infrared starting from QCD may lead
to the $O(4)$ fixed point, it may miss the fixed point and 
yield a first order transition, or, least likely, it may lead
to some as yet undiscovered {\it other} fixed point, with 
different exponents. The first order option does in fact arise
if the strange quark is lighter than a certain 
(tricritical\cite{wil,RW1,myreview}) value.
Simulations with two light and one heavier
quark done by the Columbia group\cite{columbia} suggest that
in nature, the strange quark is heavy enough that no first order 
transition occurs.  It is very important to verify this conclusion
in the next generation of lattice simulations, particularly
as it has recently been questioned.\cite{iwasaki}  A final
logical possibility which would lead to nonobservation of 
$O(4)$ exponents is that the Ginzburg region may be too small
to see the true critical behavior. Fluctuations of the order
parameter are important only close to $T_c$ --- for 
$|t|\equiv |T-T_c|/T_c < t_G$ where $t_G$, like $T_c$, is not
universal.  Outside the Ginzburg region, that is for $|t|>t_G$,
mean field theory is valid.  
If there is an appropriate small parameter in the theory, 
$t_G$ can be small.
For example,\cite{BCS} 
in ordinary, low temperature, BCS
superconductors, $t_G \sim (T_c/E_f)^4$ where the Fermi
energy $E_f$ is $10^3$ to $10^4$ times bigger than $T_c$.
As another example,\cite{kogkoc} numerical simulations
of the $2+1$ dimensional Gross-Neveu model suggest that 
$t_G$ is small in this theory.  Here, the small parameter
is $1/N$, where $N$ is the number of fermions in the theory.
In the large $N$ limit in this theory, the phase transition remains in the
Ising universality class but $t_G$ goes to zero like $1/N$.\cite{pisstep}
(Note that in QCD, increasing either $N_c$ or $N_f$ has 
much more drastic effects like making the transition first order
or changing the symmetry of the order parameter or destroying 
asymptotic freedom.)
There is no parameter in QCD, small or not so small, 
which when taken to zero
reduces $t_G$, leaving the transition otherwise unaffected. 
We can therefore expect
$t_G$ to be of order one. To exclude the
possibility that it is unexpectedly small, one needs lattice
simulations with light enough quarks that 
$m_\pi \sim T_c$.  This should suffice to 
distinguish between 
$O(4)$ and mean field exponents.}
 
We have seen that the long wavelength oscillations of the orientation of the
condensate are central to the physics of the chiral phase transition.
Although measuring critical exponents, and thus 
gaining a quantitative understanding of this physics,
is possible on the lattice, 
it is extremely unlikely that such measurements can
be done in heavy ion collision experiments.  However,
as I explain in the rest of this talk, classical long
wavelength pion oscillations may arise in a heavy ion collision,
and can leave a signature.\cite{RW2,myreview,blkrreview}

A long wavelength oscillation of the order parameter 
consists of large regions in which the chiral condensate
points in
directions other than $(\sigma,\vec 0)$.
Let us first consider an idealized
situation \cite{anselm,blaizot,bj}
in which there is a single large region in which $\phi$ 
is uniform in space and misaligned.  
Because the pion mass is nonzero due to the explicit
chiral symmetry breaking introduced by nonzero 
quark masses, in such a region of disoriented
chiral condensate the $\phi$ field would
oscillate about the $\sigma$ direction.  That is, say,
\begin{equation}
\phi = v( \cos \theta, 0,0,\sin\theta) {\rm ~with~} \theta(t) \sim
\sin ( m_\pi t) \ .
\label{idealdcc}
\end{equation}
In this idealized case, $\phi$ is 
independent of $\vec x$.  This can be thought of as an infinitely long
wavelength oscillation of the orientation of the condensate, unpolluted
by short wavelength clutter.
(In a heavy ion collision,
we will see that if long wavelength oscillations arise, they
occur superposed
with short wavelength ``noise'' and so a smooth
region of disoriented chiral condensate 
is an idealization.
Bjorken {\it et al} have suggested that circumstances closer to the
idealized case may actually arise in hadron hadron collisions.\cite{bj})

An idealized disoriented region in which the disorientation
is in the $\pi_1 - \pi_2$ plane corresponds to all charged pions
(equally positive and negative since the fields are real),
while if the disorientation is in the $\pi_3$ direction
as in (\ref{idealdcc}) it corresponds to all neutral pions
with no charged pions.
More generally if we define the ratio
\begin{equation}
R\equiv \frac{ n_{\pi^0}}{n_{\pi^0} + n_{\pi^+\pi^-}}
\label{Rdef}
\end{equation}
then each disoriented region will yield pions with some fixed $R$.
Under the assumption that all directions of disorientation are
equally probable, an ensemble of ``events'' will yield
a probability distribution for $R$ \cite{andreev,anselm,bj,RW2}
\begin{equation}
P(R) = \frac{1}{2\sqrt{R}}\ .
\label{invsqrt}
\end{equation}
Note, for example, that the probability that the neutral
pion fraction $R$ is less than 0.01 is 0.1! 
In ``events'' with, say, 100 pions, the distribution (\ref{invsqrt})
is very different than the  binomial distribution obtained
if each individual pion was independently randomly neutral or
positive or negative.  Rather than rolling the dice once per pion,
and getting an $R$ distribution looking like a Gaussian
about $R=1/3$, in a population of idealized ``events'' just described,
the dice are rolled once per event yielding the much broader, and
skewed, distribution (\ref{invsqrt}).   

Now with a hint of how long wavelength oscillations of the order parameter
could leave a signature, we return to the question of whether
they arise.  An equilibrium second order transition seems ideal, as near the
critical temperature the correlation length is infinite and oscillations
occur on arbitrarily long wavelengths.  However, because the up
and down quarks (and hence the pion) are not massless, we have
seen that the putative second order transition is smoothed out,
and the correlation length is not long compared to $1/T_c$, and
classical oscillations do not arise.  
Although we are used to thinking of nature as being close
to the chiral limit, because $m_\pi/m_\rho$ is small, it is unfortunately
not close enough for $m_\pi/T_c$ to be small.
On the 
lattice, one can envision using quark masses smaller than in nature,
and so getting closer to having a second order transition. In an experiment,
one does not have this option.

Fortunately, this is not the end of
the story.\cite{RW2}  There is no reason to assume that during a heavy ion
collision the long wavelength modes of the order parameter stay
in thermal equilibrium as the plasma expands and cools and chiral
symmetry breaking occurs, 
even if local thermal equilibrium at a high temperature is achieved
at early times. There is a 
simple non-equilibrium mechanism,
which could plausibly operate in a heavy ion collision,
which naturally leads to great amplification of long wavelength
modes, relative to what would be seen in thermal equilibrium.

Let us consider an
idealization that is in some ways opposite
to that of thermal equilibrium, namely the occurrence of a sudden
quench from high to low temperatures in which the 
$(\sigma,\vec\pi)$ fields are suddenly removed
from contact with a high temperature heat bath and
subsequently evolve according to zero temperature
equations of motion.\footnote{Quenching is an {\it ad hoc}
assumption, not expected to be realized in a heavy ion collision
in any detail, and so it is reasonable to
consider different 
dynamical assumptions as has been done in \cite{otherdynamics,othermodels}.  
The hope is that qualitative behavior observed
after a quench is representative of phenomena 
occurring in
realistic conditions in heavy ion
collisions in which the long wavelength modes are not in
equilibrium.}  
Unlike in the equilibrium case where universality was our guide, 
away from equilibrium we must make a non-universal choice
of model Lagrangian to obtain equations of motion for the
order parameter.  
The linear sigma model
\begin{equation}
{\cal L} = \int d^4 x \left\{ \frac{1}{2} \partial^\mu\phi^\alpha
\partial_\mu \phi_\alpha - \frac{\lambda}{4}\left(\phi^\alpha
\phi_\alpha - v^2\right)^2 + H\sigma \right\}\ ,
\label{linsigmodel}
\end{equation}
is a reasonable choice\cite{RW2}
although other models have also been studied.\cite{othermodels}
In \cite{RW2}, the nonlinear 
classical equations of motion derived from this
Lagrangian were solved numerically
on a lattice with spacing $a = (200 {\rm MeV})^{-1}$,
and with parameters in the potential chosen such that $f_\pi$ and
$m_\pi$ have their zero temperature values and $m_\sigma = 600{\rm MeV}$.
In the initial conditions, $\langle \phi \rangle \sim 0$ and
the orientation of $\phi$ was chosen randomly at each lattice
site, as appropriate for initial conditions 
above $T_c$ at a high enough
temperature that the correlation length
is shorter than $a$.   
We chose $\langle\phi^2\rangle^{1/2} = v/2$,
a choice which has since received some justification from the 
work of Randrup, as we discuss below.
Because of the
explicit symmetry breaking, 
$\phi$ is soon oscillating about the $\sigma$ direction
everywhere in space.  Upon Fourier transforming, one finds that
modes of each component of $\vec\pi$ with wave vector $\vec k$ 
oscillate about $\vec\pi=0$ with $\omega\sim\sqrt{k^2 + m_\pi^2}$,
as expected.  
The 
behavior of the amplitude of the low momentum pion modes 
as a function of time is striking.
The initial conditions had a white noise power spectrum, with
all modes having equal amplitudes.
At very late times, things become boring once again, as  
the system approaches an equilibrium configuration in 
which equipartition of energy holds.  
Over a wide range of intermediate times of order
$5$ to $50$ times $m_\pi^{-1}$, however, the long wavelength modes are
greatly amplified.  (This was confirmed in \cite{ggp,bialas}.)
The affected modes are 
those with $|\vec k|$ less than about 
$m_\pi$.  The longer the wavelength, the bigger the amplification; 
the longest wavelength mode not affected
by the finite size of the box has its amplitude squared 
amplified by more than a factor of $1000$ relative to that of
the short wavelength modes which were not amplified at all, and
by about a factor of $50$ relative to its value at late times in
equilibrium.  
For a visual image, think of long wavelength
ocean swells superposed with lots of short wavelength chop. 
The essential qualitative phenomenon whose cause and consequences
we now discuss is that long wavelength oscillations of the
orientation of the condensate (pion oscillations) have been excited ---
the short wavelength modes have {\it not} been damped out. 
One does not see smooth domains, and the power spectrum is not
well characterized by a single correlation length
or domain size.\cite{myreview}
This picture
is more complicated than the idealized ``smooth'' region of disoriented
chiral condensate we discussed above.

The emergence of long wavelength pion oscillations is a striking
qualitative phenomenon, for which there is a simple qualitative
explanation.\cite{RW2}  One can linearize the equations
of motion for the pion field to obtain
\begin{equation}
\frac{d^2}{dt^2} \vec\pi(\vec k,t) = -m_{\rm eff}^2(k,t)\vec\pi(\vec k,t)\ ,
\label{lineq}
\end{equation}
where the time dependent ``mass'' is given by 
\begin{equation}
m_{\rm eff}^2 \sim -\lambda v^2 + k^2 + \lambda \langle \phi^2 \rangle(t).
\label{meff}
\end{equation}
Linearizing in this way constitutes an uncontrolled truncation, but
it does yield some insight into the behavior found by numerical 
solution of the 
nonlinear equations. One sees that the non-equilibrium dynamics
leads to a time dependent $m_{\rm eff}^2$ which can be
negative for modes
with small enough $k$.  This means that these
modes will experience periods of time during which
they are unstable to exponential growth.  The longer the wavelength,
the greater the amplification.  In an equilibrium phase transition,
explicit symmetry breaking keeps the correlation length at 
$T_c$ too short to be of interest.   
This, then, is
a simple non-equilibrium mechanism which 
leads to the amplification of arbitrarily long wavelength modes
of the pion field even though the pion mass is nonzero.

I will return to analyzing the results of the simulation just
described in terms of the ratio $R$, but let me first sketch
some improvements.  The most important effect which
has been left out above is expansion. 
The late time behavior with no expansion (equilibration of the system)
is qualitatively different than that with expansion. 
Expansion causes the energy density
to drop, and the dynamics linearizes as 
the modes stop interacting at late times. Hence, at late times the 
ratios of amplitudes between different modes stops changing. This is 
the classical analogue of ``freeze out''.  The
hope in \cite{RW2} was that once expansion is included, the 
modes would freeze out with amplitude ratios as found 
at times of order $5 - 50\ m_{\pi}^{-1}$, namely with long
wavelength modes amplified.  This was in fact found in
some,\cite{AHW,cooper,lampert} but not all,\cite{bialas}
subsequent simulations which began with 
initial conditions akin to those above and which
included effects of 
expansion in several different ways. (For treatments
in other contexts which include effects of expansion, 
see \cite{otherexpansion}.)

Another goal in implementing expansion, however, is to 
relax the quench assumption.  
This has been implemented by Randrup.\cite{randrup}  He begins
with a linear sigma model configuration in thermal equilibrium
at $400\ {\rm MeV}$,\footnote{Although this is certainly less
{\it ad hoc} than the initial conditions in \cite{RW2}, it is
not yet realistic.  At high temperatures like these, at
wavelengths of order $1/T$ there are presumably degrees of
freedom other than those of the linear sigma model which are important.}
 and adds a damping term to the equations
of motion which damps energy from the system as if the system
is undergoing a three dimensional boost invariant expansion.
Initially, 
$\langle \phi \rangle$ is small, but because of the large initial 
temperature
$\langle \phi^2 \rangle$ 
is much larger than in the initial conditions of \cite{RW2}.  
Randrup finds 
an initial rapid decrease in 
$\langle \phi^2 \rangle$, during which $\langle \phi \rangle$
hardly changes and $\langle \phi^2 \rangle$ decreases to  $\sim v^2/2\,$!
Thus, the expansion takes the system to a configuration
similar to that configuration chosen in \cite{RW2} 
as an initial condition for evolution after a quench, which
in turn was supposed to model the effects of expansion.
The  subsequent evolution in Randrup's simulation
is similar to that in \cite{RW2}; 
long wavelength modes grow, 
although because of the expansion they later
freeze out as described above. Thus, Randrup's result
provides some support for some of the {\it ad hoc} choices
made in \cite{RW2}. 

A more ambitious project than the classical simulations
described so far is to include the effects of quantum
fluctuations (in addition to classical thermal fluctuations)
on the dynamics of the order parameter.  
We have seen that with the pion mass nature has dealt us,
the equilibrium correlation length is not long compared
to $1/T_c$, and this suggests that quantum effects may matter.
However, the nice thing is
that the instability discussed above is an instability towards
growth of long wavelength modes, which as they grow in amplitude 
become
{\it more} classical. In 
equilibrium, the occupation number of a mode is
$\sim T/\omega$ which is  $\sim T/m_\pi$ at long wavelengths, 
so any growth relative to equilibrium pushes the mode into
the classical regime.  Simulations
including quantum fluctuations have now been done by 
several groups,\cite{cooper,boyanovsky,lampert}
and they in fact find the same instability 
to growth of long wavelength modes found classically. However,
in quantum treatments done 
to this point it has been necessary to make 
linearizing approximations.
I argue in \cite{myreview} that 
the simulations of \cite{RW2} show that the growth of long wavelength
modes in the full nonlinear theory is in fact greater than
in the linearized theory.  Hence, it is best
to view the two treatments as complementary. The quantum treatment
can be used to study under what circumstances the instability
is present, but once the instability sets in and long wavelength
modes begin to grow, a classical treatment becomes appropriate.
Happily, the two approaches 
seem to yield approximately similar outcomes.  

Because a non-equilibrium transition is
required, the theoretical foundations of all the simulations
are shaky.  Using the linear sigma model to describe the
dynamics seems reasonable for long wavelength modes, but
cannot be justified in a controlled way. Either neglecting
quantum effects or including them in a Hartree approximation
risks missing some of the physics.  Most important of
all, we do not know what initial conditions to impose on the
long wavelength modes of the chiral order parameter at the
time when partonic language ceases to be appropriate and a
description in terms of the linear sigma model becomes
appropriate. 
Nobody can do a simulation which follows 
non-equilibrium dynamics beginning with cascading partons and
ending with
long wavelength oscillations of the condensate.
This makes it impossible 
to {\it predict} that long wavelength pion oscillations {\it will}
be amplified in a particular experimental setting.
The way to know that this phenomenon has occurred is to detect 
it, and to this we now turn.

The defining signature\footnote{Other signatures 
which have been discussed include
an excess in the total number of pions at 
low $p_T$ independent of their charge,\cite{cooper,lampert,gavin,huang} 
various effects on
pion pair correlations,\cite{corr}
and various electromagnetic effects.\cite{suzukihuang,huang} Any
of these would
be telling if seen in conjunction with large fluctuations in $R$,
but it is not clear that they would be unambiguous if seen alone.}
of the presence of long wavelength classical pion
oscillations is large fluctuations in the ratio $R$.
If a particular $\vec k$ mode ends up oscillating in the
$\pi^3$ direction, then it will become neutral pions moving
in the direction of $\vec k$. Similarly, oscillations in the
$\pi^1 - \pi^2$ plane will become charged pions. If this
were the only physics occurring, one could bin the data as 
a function of pseudorapidity $\eta$ and azimuth $\varphi$
and the bin by bin distribution of $R$ would be (\ref{invsqrt}).
The problem, of course, 
is the large background due to all the 
other pions in the event. 

What is needed is a detector which covers as much of $\eta$
and $\varphi$ as possible, which counts both charged pions
and photons, which accepts pions down to as low $p_T$ as possible
($50 {\rm ~MeV}$ or even lower; this may mean
running with reduced magnetic field), and which accepts 
photons down to $p_T\sim m_\pi/2$
or so. The detector should be segmented in $\eta$ and $\varphi$
finely enough that the number of photons and
charged pions in each segment can be accurately counted.
Where should one be in $\eta$? Central rapidity is cleanest 
for a theorist,
but at RHIC low $p_T$ photons are hard to count there,
so perhaps somewhat off central rapidity may be better.
Accurate measurement of $p_T$ and particle identification is not necessary,
since at RHIC most charged particles are pions and most
photons are from $\pi_0$'s. However, it would be very
helpful to have enough $p_T$ information to make even a 
crude cut, keeping only pions with low $p_T$, say $p_T < 2-3\, m_\pi$.
The data, then, consists of an event by event catalogue
of the positions in $(\eta,\varphi)$ of the photons and charged
pions.  From this, one can compute $R=n_\gamma/(n_\gamma+ 2 n_{\rm ch})$
in bins in $(\eta,\varphi)$.

In thinking about how to analyze real data, it is an instructive
exercise\cite{inpreparation} to go back to
the simulations of \cite{RW2} and make 
explicit how the amplification of long wavelength
modes is manifest in the number ratio $R$.
For each of the three $\vec\pi$, and for each $\vec k$ on the
momentum space lattice with spacing $2\pi/a$, define the
``particle number''
\begin{equation}
n_{\vec k}^i(t) \equiv \frac{1}{\omega}\left| i \omega \pi^i(\vec k,t)
+ \dot\pi^i(\vec k,t)\right|^2\ ,
\label{ndef}
\end{equation}
where $\pi^i(\vec k,t)$ and $\dot\pi^i(\vec k,t)$ are the
Fourier transforms of $\pi^i(\vec x,t)$ and its time derivative.
If we had included expansion, had waited until the system linearized, and
if we assumed that the classical fields correspond to quantum mechanical
coherent states, these would be the mean particle numbers per mode
at late time. Although the $n$'s should really be viewed just
as time dependent quantities defined by (\ref{ndef}),
I will refer to them as numbers henceforth.
The lattice has $64^3$
points, so we now have the number of neutral pions and the total
number of pions at $64^3$ points in momentum space as a function of time.  
I now make a cut, keeping only those modes with $|\vec k|<300{\rm ~MeV}$.
For each remaining $\vec k$, I compute the polar angle $\theta$
and the azimuthal angle $\varphi$ describing the direction in momentum
space.  (In an experiment, one would use $\eta$. Here, without
expansion, $\theta$ is the right variable.)
I then divide momentum space into $10^\circ\times 10^\circ$
bins in $(\theta,\varphi)$, and discard the bins
within $30^\circ$ of the north and south poles, as they each
contain too few modes.  For each of the remaining bins, I add up the number
of neutral pions in all the modes with $\vec k$'s in the bin, do
the same for the total number of pions, and compute $R$.  

The results are shown in Figure 1, at two different times.  
\begin{figure}[t]
\vspace{-.7in}
\centerline{
\epsfysize=5.0in
\epsfbox{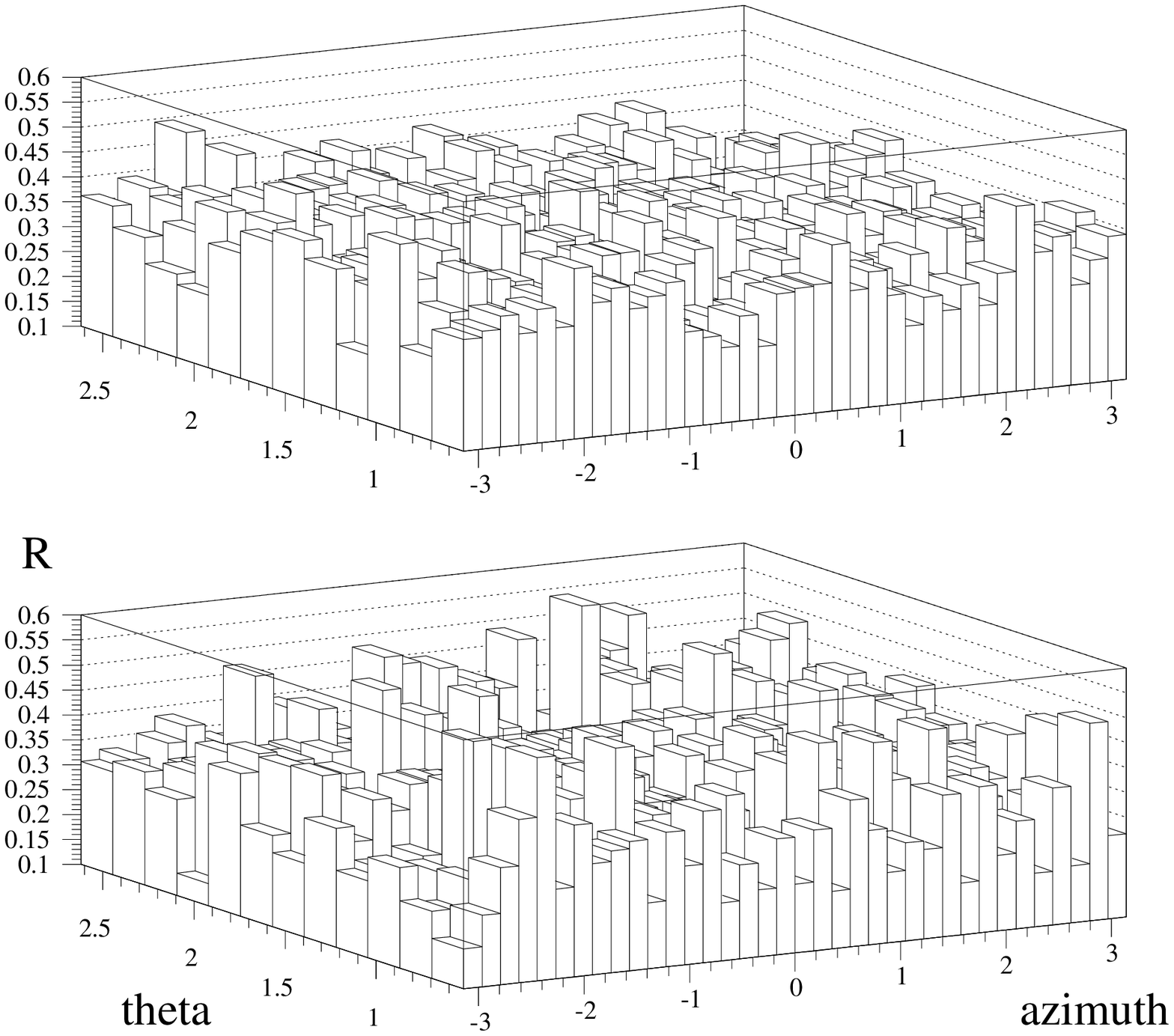}
}
\vspace{-0.95in}
\begin{center}
\begin{minipage}{13cm}
\baselineskip=12pt
{\begin{small}
Fig.~1. R as a function of angle in momentum space, binned
in $10^\circ\times 10^\circ$ bins. The top panel is at $t=160 {\rm ~fm}$
when nothing interesting is going on, and illustrates the magnitude
of ``ordinary'' fluctuations.  The bottom panel is at $t=22 {\rm ~fm}$,
a time when long wavelength modes have been amplified.\end{small}}
\end{minipage}
\end{center}
\vspace{-0.4in}
\end{figure}
\noindent
The top
panel is at a very late time, when the system has equilibrated and
nothing interesting is going on.  The bottom panel is
at an intermediate time when 
the long wavelength modes have large amplitudes.  It is quite clear
in the figures that the $R$ distribution is much broader in the
lower panel.  There are more high spots and low spots.  Figure
2 shows histograms of the distribution of $R$ values in the 432
bins in Figure 1.  
\begin{figure}[t]
\vspace{-.7in}
\centerline{
\epsfysize=5.0in
\epsfbox{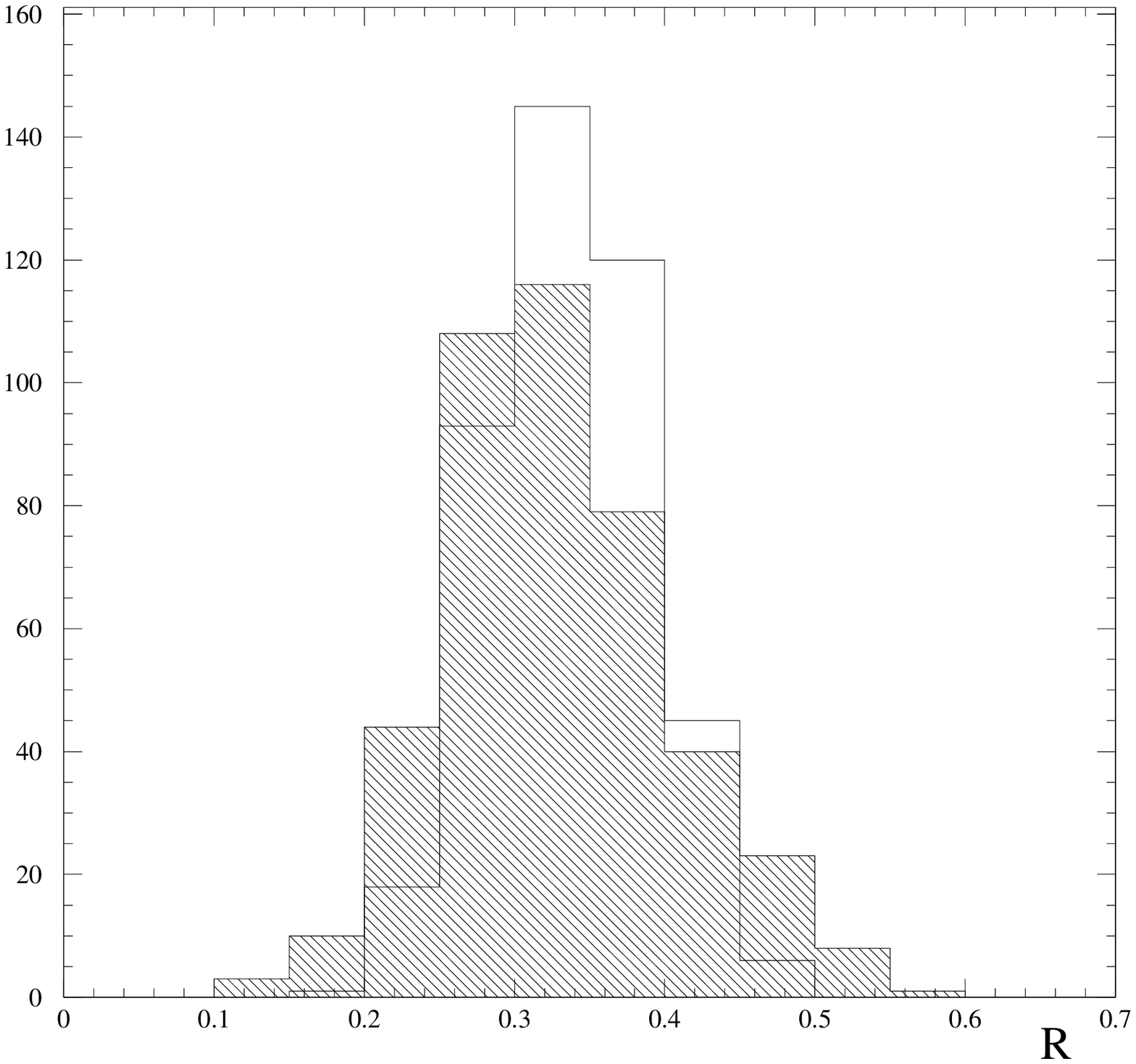}
}
\vspace{-0.9in}
\begin{center}
\begin{minipage}{13cm}
\baselineskip=12pt
{\begin{small}
Fig.~2. Histograms of the distributions of $R$ in Figure 1.
The shaded histogram corresponds to the lower panel in
Figure 1, and the histogram drawn simply as a  
line corresponds to the upper panel.\end{small}}
\end{minipage}
\end{center}
\vspace{-0.4in}
\end{figure}
\noindent
The unshaded histogram corresponds to the
top panel, and
the shaded histogram corresponds to the bottom panel, in which
long wavelength oscillations are present.  We see that the effect
of the long wavelength modes is to broaden the distribution (look
at the tails),
and skew it to the left (look at the peaks).
This is exactly what one would expect from an admixture of a
$1/\sqrt{R}$ distribution.  Note that if the $300{\rm ~MeV}$
cut on $|\vec k|$ is pushed too high, the short wavelength
modes swamp the long wavelength modes and both distributions
look the same. If the cut is too low, there are very few modes
per bin, and both distributions become wider. Eventually, if
the cut is too low (or if the bin size 
is too small) and there is only one mode per bin, then both
distributions are $1/\sqrt{R}$.  Increasing the bin size
above $10^\circ \times 10^\circ$
narrows both distributions by the same factor.
The central limit theorem ``acts'' on both distributions
as the bin size is increased, but the presence of 
large amplitude long wavelength modes ``delays its action''.\cite{wavelet}
The choice
of cuts and bin sizes in Figs. 1 and 2 
was driven by the background (short wavelength modes)
and statistical fluctuations (arising if there are too few modes per bin)
in the simulation in \cite{RW2}. 
In analyzing real data in an analogous fashion, 
such choices
will be driven by the background (high $p_T$ pions)
and statistical fluctuations (arising if there are
too few pions and photons per bin) and the fact that one
counts photons rather than $\pi^0$'s.
Hence, the choices
I have found convenient need not be the ones which are 
convenient in analyzing real data.

With real data, one should
try bins of all sizes in both $\eta$ and $\varphi$, ranging
from bins which on average contain a few tens of pions, to
bins in rapidity covering $2\pi$ in $\varphi$, to
``bins'' which consist of whole events, and construct
$R$ distributions as above for each bin size.
(The data should also be analyzed using the wavelet formalism
of \cite{wavelet}.)
It is also very important to look for singular
bins far out on the tails of the various distributions.
The simulations of \cite{AHW} suggest that events of this sort 
may occur, although they are presumably rare.
Broadened (and perhaps skewed) $R$ distributions 
may be more common. Both should be looked for as functions
of increasing beam energy and projectile size, and decreasing
impact parameter.

To what should one compare an $R$ distribution obtained from data?
One can use an $R$ distribution obtained from
an event generator  in which pions are independently randomly
neutral, positive, or negative.  If there is no signal in the
data, this should fit very well, because any badness of fit
in each individual multiplicity distribution should cancel
in the ratio $R$.  However, it may be
preferable to do the analysis without reference to an
event generator.  Steinberg\cite{WA98} has suggested creating a data
set by rotating the charged pions in each event
in azimuth relative to the photons, say by $90^\circ$.
One would
look for broadening and skewing of the histograms from the
real events relative to those from the rotated events.

Although there are fewer complications
at RHIC than at lower energies, given the lack of theoretical
certainty it is certainly wise to look today.
WA98 at CERN 
is in the process of analyzing their 
event by event charged particle
and photon 
data.\cite{WA98}
They are looking for, and so far have not found,\cite{WA98}
singular events which behave as if
more than $\sim 30\%$ of all the pions in the event 
come from a single idealized region of disoriented
chiral condensate.  
They are in the process of binning
the data in $(\eta,\varphi)$ and looking for fluctuations
in $R$ using various bin sizes.  Even before they finish their
analyses, we can learn useful lessons for future experiments.
Being able to cut on $p_T$, and get rid of the high momentum
pions, would really help, as would having a charged particle
veto in front of their photon multiplicity detector.
NA49 is planning to search for fluctuations in the number of charged
pions only, as they do not 
count photons.\cite{roland}\footnote{MINIMAX is looking for 
regions of disorientated chiral condensate at high 
rapidity in hadron-hadron
collisions at Fermilab. The dynamics in these events is
surely different from that we are discussing, but 
the signatures are similar and therefore
their analysis techniques\cite{minimaxanal}
should prove useful when generalized for use in larger acceptance
experiments.}  
PHENIX, PHOBOS, and
STAR all plan to search for unusual fluctuations in $R$ at RHIC.

The conclusions from theory for experiment are qualitative.
Consider a heavy ion collision which is energetic enough that
there is a central rapidity region of high energy density
and low baryon number. If, as this region cools through
the chiral transition, long wavelength modes do not
stay in equilibrium, then there is a robust mechanism which
can lead to the amplification of 
long wavelength pion oscillations. This cannot happen in thermal
equilibrium.  One should not take the details of any of the
theoretical simulations too seriously.  All make dynamical
assumptions and all rely on guesses for their initial conditions,
since nobody can go from initial parton dynamics to the dynamics of
the chiral order parameter at long wavelengths.  What one should
take seriously is that the place to look for signatures of 
the dynamics of the chiral order parameter during the phase
transition is the low $p_T$ pions.  We have discussed one
effect, leading to such a signature, but perhaps the 
best outcome would be the detection of completely unexpected
phenomena at low $p_T$. 
Long wavelength oscillations
of the condensate would be observed by detecting unusual fluctuations
in $R$, the number ratio of neutral pions, event by event 
as a function of 
rapidity and azimuth.  One should look both for broadened and
skewed $R$ distributions, and for singular events with particularly
large fluctuations.  
Experimentalists are trying to 
disorder the chiral condensate and then detect its 
disoriented oscillations, as it returns toward its 
ordered ground state. They will continue
at RHIC and the LHC, where sufficient energy should
be available. Detection of unusual fluctuations in $R$ in a
heavy ion collision would be a dramatic and definitive signature
of an out of equilibrium chiral transition. 
The ball is in the experimentalists' court, and we wish them
well.

{\bf Acknowledgments} --- I am grateful to the organizers
for inviting me to Hirschegg; the meeting was enjoyable and productive.
I would like to thank everyone 
at the Trento conference on Disoriented Chiral Condensates.
Many discussions I had there influenced this work. 
This work was supported in part by the Sherman Fairchild
Foundation and by the Department of Energy under Grant No.
DE-FG03-92-ER40701.


\begin{thebibliography}{99}
\itemsep=0cm
\bibitem{asympfree}
D. J. Gross and F. Wilczek, Phys. Rev. Lett. {\bf 30} (1973) 1343;\\
H. D. Politzer, {\it ibid}. 1346.
\bibitem{collins}
J. C. Collins and M. J. Perry, Phys. Rev. Lett. {\bf 34} (1975) 1353.
\bibitem{svetitsky}
B. Svetitsky and L. G. Yaffe, Nucl. Phys. {\bf B210} (1982) 423;\\
B. Svetitsky, Phys. Reports {\bf 132} (1986) 1.
\bibitem{pureglue}
For a review, see A. Ukawa, Nucl. Phy., {\bf B17} Proc.
Suppl. (1990) 118.
\bibitem{latt2ndorder}
S. Gottlieb {\it et al.}, Phys. Rev. Lett. {\bf 59} (1987) 1513;
{\bf D35} (1987) 3972; {\bf D41} (1990) 622; {\bf D47} (1993) 315;
M. Fukugita {\it et al.}, Phys. Rev. Lett. {\bf 65} (1990) 816; 
Phys. Rev. {\bf D42} (1990) 2936;
F. R. Brown {\it et al.}, Phys. Rev. Lett. {\bf 65} (1990) 2491;
F. Karsch, Phys. Rev. {\bf D49} (1993) 3791; F. Karsch and E.
Laermann, Phys. Rev. {\bf D50} (1994) 6954;
C. Bernard {\it et al.}, Phys. Rev. {\bf D45} (1992) 3854;
Phys. Rev. {\bf D54} (1996) 4585; 
G. Boyd {\it et al.}, hep-lat/9607046; 
Y. Iwasaki {\it et al.}, Phys. Rev. Lett. {\bf 78} (1997) 179;
A. Ukawa, hep-lat/9612011.
\bibitem{MILC}
C. Bernard {\it et al.}, hep-lat/9608026; hep-lat/9612025.
\bibitem{piswil}
R. Pisarski and F. Wilczek, Phys. Rev. {\bf D29} (1984) 338.
\bibitem{wil}
F. Wilczek, Int. J. Mod. Phys. {\bf A7} (1992) 3911.
\bibitem{RW1}
K. Rajagopal and F. Wilczek, Nucl. Phys. {\bf B399} (1993) 395.
\bibitem{myreview}
K. Rajagopal, in Quark Gluon Plasma 2, R. Hwa, ed. (1995) 484, hep-ph/9504310.
\bibitem{detar}
C. Detar, in Quark Gluon Plasma 2, R. Hwa, ed. (1995) 1.
\bibitem{toussaint}
D. Toussaint, Phys. Rev. {\bf D55} (1997) 362.
\bibitem{bmn}
G. Baker, B. Nickel and D. Meiron, Phys. Rev. {\bf B17} (1978) 1365;
and an unpublished University of Guelph report (1977).
\bibitem{kk}
K. Kanaya and S. Kaya, Phys. Rev. {\bf D51} (1995) 2404.
\bibitem{columbia}
F. R. Brown {\it et al.} in \cite{latt2ndorder}.
\bibitem{iwasaki}
Y. Iwasaki {\it et al.}, Phys. Rev. {\bf D54} (1996) 7010.
\bibitem{BCS}
E. M. Lifschitz and L. P. Pitaevskii, Statistical Physics Part 2,
(Pergamon, Oxford, 1980).
\bibitem{kogkoc}
A. Kocic and J. Kogut, Phys. Rev. Lett. {\bf 74} (1995) 3109.
\bibitem{pisstep} 
R. Pisarski and M. Stephanov, private communication.
\bibitem{RW2}
K. Rajagopal and F. Wilczek, Nucl. Phys. {\bf B404} (1993) 577.
\bibitem{blkrreview}
For a beautiful review, see J.-P. Blaizot and A. Krzywicki, 
Acta Phys. Polon. {\bf 27} (1996) 1687.
\bibitem{anselm}
A. Anselm, Phys. Lett. {\bf 217B} (1988) 169;\\ A. Anselm and
M. G. Ryskin, Phys. Lett. {\bf 266B} (1991) 482.
\bibitem{blaizot}
J.-P. Blaizot and A. Krzywicki, Phys. Rev. {\bf D46} (1992) 246.
\bibitem{bj}
J. D. Bjorken, Int. J. Mod. Phys. {\bf A7} (1992) 4189;
Acta Phys. Pol. {\bf B23} (1992) 561; K. L. Kowalski and C. C.
Taylor, hep-ph/9211282; J. D. Bjorken, K. L. Kowalski and C. C.
Taylor, hep-ph/9309235.
\bibitem{andreev} 
I. V. Andreev, JETP Lett. {\bf 33} (1981) 367.
\bibitem{otherdynamics}
A. Krzywicki, Phys. Rev. {\bf D48} (1993) 5190;
S. Gavin and B. M\"uller, Phys. Lett. {\bf B329} (1994) 486;
S. Mrowczynski and B. M\"uller, Phys. Lett. {\bf B363} (1995) 1.
\bibitem{othermodels}
P. F. Bedaque and A. Das, Mod. Phys. Lett. {\bf A8} (1993) 3151;
A. Barducci {\it et al.}, Phys. Lett. {\bf B369} (1996) 23;
J. I. Kapusta and A. P. Vischer, nucl-th/9605023;
A. Abada and M. C. Birse, hep-ph/9612231.
\bibitem{ggp}
S. Gavin, A. Gocksch and R. Pisarski,
Phys. Rev. Lett. {\bf 72} (1994) 2143. 
\bibitem{bialas}
A. Bialas, W. Czyz and M. Gmyrek, Phys. Rev. {\bf D51} (1995) 3739.
\bibitem{AHW}
Z. Huang and X. Wang, Phys. Rev. {\bf D49} (1994) 442;\\
M. Asakawa, Z. Huang, and X. Wang, Phys. Rev. Lett. {\bf 74} (1995) 3126.
\bibitem{cooper}
F. Cooper {\it et al.} Phys. Rev. {\bf D50} (1994) 2848;
{\bf D51} (1995) 2377; {\bf C54} (1996) 3298; 
Y. Kluger, hep-ph/9405279; hep-ph/9408286; hep-ph/9503205.
\bibitem{lampert}
M. A. Lampert {\it et al.}, Phys. Rev. {\bf D54} (1996) 2213.
\bibitem{otherexpansion}
J.-P. Blaizot and A. Krzywicki, Phys. Rev. {\bf D50} (1994) 442;
S. Yu. Khlebnikov, Mod. Phys. Lett. {\bf A8} (1993) 3971;
Z. Huang and M. Suzuki, Phys. Rev. {\bf D53} (1996) 891;
H. Davoudiasl, hep-ph/9611263.
\bibitem{randrup}
J. Randrup, Phys. Rev. {\bf D55} (1997) 1188; Phys. Rev. Lett. {\bf 77}
(1996) 1226; hep-ph/9612453.
\bibitem{boyanovsky}
D. Boyanovsky {\it et al.}, Phys. Rev. {\bf D51} (1995) 734; 
Phys. Rev. {\bf D54} (1996) 1748.
\bibitem{gavin}
S. Gavin, Nucl. Phys. {\bf A590} (1995) 163; hep-ph/9407368;
\bibitem{huang}
Z. Huang, hep-ph/9501366.
\bibitem{corr}
J. P. Blaizot and D. Diakonov, Phys. Lett. {\bf B315} (1993) 226;\\
C. Greiner, C. Gong and B. M\"uller, Phys. Lett. {\bf B316} (1993) 226.
\bibitem{suzukihuang}
Z. Huang, M. Suzuki, and X. Wang, Phys. Rev. {\bf D50} (1994) 2277;
{\bf D52} (1995) 2610; D. Boyanovsky {\it et al.}, hep-ph/9701360.
\bibitem{inpreparation}
K. Rajagopal, work in preparation sketched here.
\bibitem{wavelet}
Z. Huang {\it et al.}, Phys. Rev. {\bf D54} (1996) 750.
\bibitem{WA98}
T. Peitzmann, talk at this meeting; talks by T. Nayak, 
P. Steinberg, J. Urbahn and
B. Wyslouch at Trento meeting on Disoriented Chiral Condensates, 1996.
\bibitem{roland}
G. Roland, talk at this meeting.
\bibitem{minimaxanal}
T. C. Brooks {\it et al.}, hep-ph/9609375; J. D. Bjorken {\it et al.},
hep-ph/9610379.

\end{thebibliography}
\end{document}